\documentclass[aps,prl,preprint,superscriptaddress,showpacs]{revtex4}

\usepackage{bm}
\usepackage{amsmath}
\usepackage{amsfonts}
\usepackage{amssymb}
\usepackage{longtable}
\usepackage{epsf}
\usepackage[dvips]{epsfig,color}
\usepackage{version}
\usepackage{graphicx}
\usepackage{psfrag}

\newcommand{\be}{\begin{equation}}
\newcommand{\ee}{\end{equation}}

\begin{document}


\title{Short-time diffusion behavior of Brownian particles in confining potentials}

\author{Daniel Schneider}
\author{Rustem Valiullin}
\affiliation{Faculty of Physics and Earth Sciences, University of Leipzig, 04103
Leipzig, Germany}

\author{Nail Fatkullin}
\affiliation{Institute of Physics, Kazan Federal University, Kazan 420008, Tatarstan, Russia}


\begin{abstract}
  Diffusion behavior of Brownian particles in confined spaces was studied for the displacements notably shorter than the confinement size. The confinements, resembling structure of porous solids, were modeled using a spatially-varying potential field with an infinitely large potential representing the solid part and zero potential in the void space. Between them, a smooth transient mimicking the interaction potential of the tracer molecules with the pore walls was applied. The respective Smoluchowski equation describing diffusion of tracer particles in the thus created force field was solved under certain approximations allowing for a general analytical solution. The time-depended diffusion coefficient obtained was found to agree with that obtained earlier using the Fick's diffusion equation, but with a different numerical constant. Numerical solution of the Smoluchowski equation with selected interaction potentials was used to clarify the origin of this discrepancy. The conditions under which the solutions obtained within these two approaches converge were established.
\end{abstract}




\maketitle

\section{Introduction}

Diffusion, or Brownian motion of small matter constitutes, is an ubiquitous phenomenon in nature and often plays a key role in various practical applications. Under equilibrium conditions and in spatially-homogeneous and isotropic systems, the Green's function, i.e. the probability density that a particle displaces by a vector $\mathbf{r}$ within a time interval $t$, is a well-known Gaussian function \cite{Gillespie2012}. As a consequence, the mean square displacements (MSD) $\langle \mathbf{r}^{2}(t) \rangle$ are found to grow linearly with the observation time. This regime is often referred to in the literature as normal diffusion. In many cases, however, diffusion takes place in inhomogeneous media, with the most prominent examples being water transport in biological cells or fluid transport in porous solids. Under these conditions, some part of the space  becomes inaccessible for the diffusing species, causing the process of diffusion to deviate from the normal patterns on the length scale comparable to a characteristic size $\xi$ of the structural inhomogeneities. Concerning porous solids, $\xi$ may be regarded to be of the order of an average pore size. Quite generally, over these particular length and corresponding to it time scales MSD grow more slowly than in the case of normal diffusion, i.e. $\langle \mathbf{r}^{2}(t)\rangle \propto t^{k}$ with $0<k<1$ ($k$ may also be a function of time). This regime is commonly referred to as sub- or anomalous diffusion. For permeable media, with the porosities above the percolation threshold, the normal diffusion behavior is recovered for displacements notably exceeding $\xi$ \cite{Bouchaud1990, Bunde1996}. The time-dependent MSD contains thus essential information about the pore space, rendering their measurement a powerful tool of structural analysis. The developments of theoretical models allowing for the extraction of structural information encoded in MSDs become thus an important task \cite{Kac1966, Grebenkov2007}. Better understanding of the structure-dynamics relationships may further serve as a basis for predicting molecular transport in practical applications involving porous solids, including oil recovery, water infiltration, and cellular transport \cite{Cussler2009}.

A major impact on the establishment of the structure-dynamics relationships for fluids in different classes of porous solids has been provided by the diffusion measurements using nuclear magnetic resonance (NMR). The pulsed field gradient technique of this method (PFG NMR) supplies the most direct way to non-perturbatively and non-invasively trace MSDs \cite{Kimmich1997, Price2009, Callaghan2011}. It is essentially based on encoding and subsequent decoding of the nuclear spin positions via their precessional frequencies coming forth in a magnetic field. For this purpose, short pulses of the magnetic field gradients are typically used. The two encoding and decoding pulses are separated by a well-defined time interval $t$ (also referred to as diffusion or observation time) during which the molecules diffuse. In PFG NMR, the time $t$ can be varied in a wide range from about one to several thousands of milliseconds. This allows to cover the most interesting length scale from tens to thousands of nanometers, which overlaps with the characteristic length scales typical for variety of industrially- and biologically-relevant heterogeneous media.

The signal intensity $S$ measured using PFG NMR turns out to be a spatial Fourier transform of the Green's function. Thus, if the molecules during the time interval $t$ perform purely stochastic, translational motion, $S$ has a simple, exponential form

\begin{equation}
S = S_{0} \exp \left \{ - \frac{1}{2} q^{2} \langle \mathbf{r}^{2}(t) \rangle \right \},
\label{SEA}
\end{equation}

where $q$ is the wave number defined by the experimentally-controlled parameters and $S_{0}$ is the signal intensity obtained without applying the magnetic field gradients ($q=0$). Alternatively, Eq.~(\ref{SEA}) may also be considered as a first term in the expansion if the Green's function deviates from the Gaussian shape \cite{Callaghan2011}.

It is more convenient to define the effective diffusion coefficient $D_{e}$ via the Einstein equation,

\begin{equation}
D_{e}(t) = \langle \mathbf{r}^{2}(t) \rangle /6t,
\label{DEFF}
\end{equation}

instead of using the experimentally measured quantity $\langle \mathbf{r}^{2}(t) \rangle$. For the molecular displacements $\sqrt{ \langle \mathbf{r}^{2}(t) \rangle } << \xi $, $D_{e}$ coincides with the bulk self-diffusivity $D_{0}$ of the fluid under study. In porous materials, however, $D_{e}$ is found to decrease with increasing time for the diffusion times $t \sim \xi^{2}/D_{0}$. As mentioned earlier, the time dependence results as a consequence of the reflecting boundary conditions for the molecular fluxes at the pore walls. Hence, one may expect $D_{e}$ to scale as a function of the surface area. In the context of the PFG NMR diffusion measurements for fluids confined in porous solids, a notable progress in understanding of the relationships between the structural parameters of porous media and the time-dependent diffusivities was put forward in Refs. \cite{Mitra1992, Mitra1993, Latour1993}. In particular, by considering the classical Fick's diffusion equation in a space with the dimension $d$ and with the boundary condition posing zero gradient of the Green's function at the pore walls it was shown that, for sufficiently short diffusion times, $D_{e}$ is indeed determined by the surface-to-pore volume ratio $S/V_{p}$ of the pore space \cite{Mitra1992}:

\begin{equation}
D_{e}(t) = D_{0} \left( 1- \frac{4}{3d} \frac{S}{V_{p}}  \sqrt{\frac{D_{0}t}{\pi}} + O(D_{0}t) \right).
\label{STD}
\end{equation}

This result has further been confirmed in a number of the subsequent theoretical works (see, e.g., \cite{Grebenkov2007, Novikov2011}). Eq.~(\ref{STD}) has provided a convenient route for determining $S/V_{p}$ for porous materials with different geometries of their pore spaces by measuring the time-dependent diffusivity using PFG NMR \cite{Hurlimann1994, Sorland1997, Gjerdaker1999, Johns2001, Butler2002, Szutkowski2002, Miller2007, Bogdan2008}. In turn, these measurements could serve for validation of the theoretical predictions. However, the data obtained experimentally using PFG NMR can be affected by the surface relaxivities, which act as a sink for the nuclear magnetization. Hence the boundary conditions become altered precluding straightforward data analysis.

While Eq.~\ref{STD} was obtained by solving the diffusion equation with the specific boundary conditions imposed at the pore walls, an alternative to it approach can be based on the analysis on the Smoluchowski diffusion equation \cite{Doi1986}. The difference between these two approaches may be traced back to the corresponding Langevin equations. If the Fick's equation is obtained by considering only the stochastic noise term in the Langevin equation, the latter for the Smoluchowski equation contains an additional external force field $\mathbf{f}(\mathbf{r})$. The porous solids can be modeled as composed of two sub-spaces with zero and infinitely large potentials $U(\mathbf{r})$ with a smooth transition between them described by a certain interaction potential between the particles and the pore wall. Close to the pore walls the particles experience a potential change, i.e. they are subject to a force $\mathbf{f}(\mathbf{r})=-\partial U(\mathbf{r})/\partial \mathbf{r}$. Exactly this force field, which appears in the Smoluchowski equation, replaces effectively the boundary conditions in the classical diffusion equation approach.

Describing diffusive dynamics of confined fluids by using the Smoluchowski diffusion equation may be beneficial under conditions of either very tight confinements, such as for fluids in nanoscale pores of a few nanometers diameter \cite{Feng2007}, or of spatially-extended confining potentials, such as for particles trapped in optical traps \cite{Volpe2013}. In the context of the time-dependent diffusivities as obtained using PFG NMR, this approach has been partly addressed in the literature and its potentials have been highlighted \cite{Chang1975, Fatkullin1990, Maklakov1992, Valiullin2001}. In this work, we focused specifically on the short-time diffusion behavior, as exemplified by Eq.~(\ref{STD}) resulting from the Fick's diffusion equation. Because with the Smoluchowski diffusion equation the interaction of molecules with the solid can be described on a more fundamental level, the main goal of this work was to analyze the short-time diffusion behavior within this approach. In particular, we aimed at establishing the conditions under which Eq.~(\ref{STD}) can also be obtained in this way and to compare the approximations needed to be done within the two routes.

\section{Theoretical analysis}

In this work, the impact of the solid on the confined molecules was described by means of an effective potential $U(\mathbf{r})$. Within the solid, which is impermeable for the Brownian particles, $U(\mathbf{r})\rightarrow \infty$, while in the pore space $U(\mathbf{r}) \rightarrow 0$. At the pore walls, the potential falls to zero within a surface layer of the thickness $a_{0}$, which can be of the order of a few molecular sizes. The transition is considered to be smooth. In addition, only stationary fields $U(\mathbf{r})$ are considered. This implies that the particle mass is considered to be negligibly small as compared to the mass of the solid, thus the motion of the latter can be ignored.

The probability density $W(\mathbf{r};t)$ for the molecular displacements, namely the Green's function, satisfies the Smoluchowski equation

\begin{equation}
\frac{ \partial }{ \partial t} W(\mathbf{r};t)  = D_{0} \frac{\partial}{\partial \mathbf{r}} \left [ \frac{\partial}{\partial \mathbf{r}} - \frac{1}{kT} \mathbf{f}(\mathbf{r}) \right ] W(\mathbf{r};t),
\label{SE}
\end{equation}

where $D_{0}$ is the diffusivity in free fluid, $\mathbf{f}=-\partial U(\mathbf{r})/\partial \mathbf{r}$ is the force induced by the interaction potential near the pore walls, $k$ is the Boltzmann constant, and $T$ is temperature. Because the differential operator on the right hand side of Eq.~(\ref{SE}) is not self-adjoint, the mathematical treatment becomes increasingly complex. To simplify the problem, a new function $\Psi(\mathbf{r},t)$, which is related to $W(\mathbf{r},t)$ as

\begin{equation}
\Psi(\mathbf{r},t) = W(\mathbf{r},t) \exp \left \{ \frac{U(\mathbf{r})}{2kT} \right \},
\label{PSI}
\end{equation}

can be introduced. With Eq.~(\ref{PSI}), the Smoluchowski equation Eq.~(\ref{SE}) becomes

\begin{eqnarray}
\frac{ \partial }{ \partial \tau} \Psi(\mathbf{r};t) &= \Delta \Psi (\mathbf{r};t) \\
& + \frac{1}{2} \left [ \Delta \tilde{U}(\mathbf{r}) - \frac{1}{2} (\nabla \tilde{U}(\mathbf{r}))^{2} \right ] \Psi(\mathbf{r};t),
\label{SE2}
\end{eqnarray}

where $\tau = D_{0}t$ and $\tilde{U}(\mathbf{r}) = U(\mathbf{r})/kT$.

If the Green's function is known, MSDs can readily be obtained:

\begin{equation}
\begin{aligned}
\langle \mathbf{r}^{2}(t) \rangle = & \frac{1}{Z} \int\int d^{3}\mathbf{r}_{1} d^{3}\mathbf{r}_{0} (\mathbf{r}_{1}-\mathbf{r}_{0})^{2}  W(\mathbf{r}_{1},\mathbf{r}_{0};t) \\
& \times \exp \left \{- \frac{U(\mathbf{r}_{0})}{kT} \right \},
\end{aligned}
\label{MSD}
\end{equation}

where the statistical integral $Z$ is given by

\begin{equation}
Z = \int d^{3}\mathbf{r} \exp \left \{ - U(\mathbf{r})/kT \right \}.
\label{SSUM}
\end{equation}

The integration on the right hand side of Eq.~(\ref{SSUM}) yields a straightforward result $Z=V_{p}$. The Boltzmann factor in the integral on the right hand side of Eq.~(\ref{MSD}) takes into account the equilibrium distribution of the particles at $t=0$. The effective diffusivity $D_{e}$ is related to the MSDs via Eq.~(\ref{DEFF}).

It has earlier been shown that Eq.~(\ref{SE}) can be solved exactly to yield the effective diffusivity $D_{e}$ \cite{Fatkullin1990, Maklakov1992}:

\begin{equation}
D_{e}(t) = D_{0} \left [ 1 - \frac{D_{0}}{3t (kT)^{2}} \int_{0}^{t} \int_{0}^{t_{1}} dt_{1} dt_{2} \langle \mathbf{f}(t_{2})\mathbf{f}(0) \rangle \right ],
\label{DE_AF}
\end{equation}

where the force-force autocorrelation function $\langle \mathbf{f}(t)\mathbf{f}(0) \rangle$ is defined as

\begin{equation}
\begin{aligned}
\langle \mathbf{f}(t)\mathbf{f}(0) \rangle = & \frac{1}{Z} \int\int d^{3}\mathbf{r}_{1} d^{3}\mathbf{r}_{0} W(\mathbf{r}_{1},\mathbf{r}_{0};t) \\
& \times \exp \left \{ -\frac{U(\mathbf{r}_{0})}{kT} \right \}  \frac{\partial}{\partial \mathbf{r}_{1}} U(\mathbf{r}_{1}) \frac{\partial}{\partial \mathbf{r}_{0}} U(\mathbf{r}_{0}).
\end{aligned}
\label{FF_AF}
\end{equation}

It is interesting to emphasize here that, according to Eq.~(\ref{DE_AF}), the initial rate of change of the diffusivity $D_{e}$ is determined by only the molecules which initially were located within the layer of the thickness $a_{0}$ close to the pore walls. Otherwise, $\mathbf{f}(0)=0$ and the integral in Eq.~(\ref{DE_AF}) is zero as well for all times. Thus, because at initial times these particles are found to be under a continuous action of the force induced by the interaction potential, one may anticipate that the rate $d D_{e}/ dt$ may deviate from the predictions of the classical diffusion equations, where there are no net forces. On the other hand, the time-dependence is determined not only by the force acting at initial times, but also by the statistics of the surface re-visits. In particular, by assuming that the forces act only instantaneously during the reflection instances (elastic collisions) and by assuming that the re-visits statistics is described by the Green's function for a half-space, Eq.~(\ref{STD}) has been shown to follow from Eq.~(\ref{FF_AF}) \cite{Valiullin2001}. Whether this finding is also applicable for arbitrary potential fields will be in the focus of the subsequent analysis.

By substituting $W(\mathbf{r}_{1},\mathbf{r}_{0};t)$ in Eq.~(\ref{FF_AF}) with

\begin{equation}
W(\mathbf{r}_{1},\mathbf{r}_{0};t)= \exp \left \{- \frac{U(\mathbf{r}_{1})}{2kT} \right \} \Psi(\mathbf{r}_{1},\mathbf{r}_{0};t),
\label{PSI_1}
\end{equation}

one finds

\begin{equation}
\begin{aligned}
\langle \mathbf{f}(t)\mathbf{f}(0) \rangle = & \frac{(2kT)^{2}}{Z} \int\int d^{3}\mathbf{r}_{1} d^{3}\mathbf{r}_{0} \frac{\partial}{\partial \mathbf{r}_{1}} \exp \left \{ -\frac{U(\mathbf{r}_{1})}{2kT} \right \} \\
& \times \Psi(\mathbf{r}_{1},\mathbf{r}_{0};t) \frac{\partial}{\partial \mathbf{r}_{0}} \exp \left \{ -\frac{U(\mathbf{r}_{0})}{2kT} \right \}.
\end{aligned}
\label{FF_AF2}
\end{equation}

Notably, $\Psi(\mathbf{r}_{1},\mathbf{r}_{0};t)$ satisfies Eq.~(\ref{SE2}). The formal solution of Eq.~(\ref{SE2}) is

\begin{equation}
\Psi(\mathbf{r}_{1},\mathbf{r}_{0};t)= \exp \left \{ \tau \hat{L}_{1} \right \} \Psi(\mathbf{r}_{1},\mathbf{r}_{0};0),
\label{PSI_SOL_1}
\end{equation}

where the Liouville operator $\hat{L}_{1}$ is given by

\begin{equation}
\hat{L}_{1} = \Delta_{1} + \frac{1}{2} \left [ \Delta_{1} \tilde{U}(\mathbf{r}_{1}) - \frac{1}{2} (\nabla_{1} \tilde{U}(\mathbf{r}_{1}))^{2} \right ].
\label{L1}
\end{equation}

The subscript '1' in Eq.~(\ref{L1}) implies that the differentiation is performed over the final position coordinates $\mathbf{r}_{1}$. By noting that $W(\mathbf{r}_{1},\mathbf{r}_{0};0)=\delta(\mathbf{r}_{1}-\mathbf{r}_{0})$, where $\delta(x)$ is the Dirac delta function, Eq.~(\ref{FF_AF2}) becomes

\begin{equation}
\begin{aligned}
\langle \mathbf{f}(t)\mathbf{f}(0) \rangle = \frac{(2kT)^{2}}{Z} \int\int d^{3}\mathbf{r}_{1} d^{3}\mathbf{r}_{0} \left ( \frac{\partial}{\partial \mathbf{r}_{1}} e^{-\frac{U(\mathbf{r}_{1})}{2kT}}  \frac{\partial}{\partial \mathbf{r}_{0}} e^{-\frac{U(\mathbf{r}_{0})}{2kT}} \right ) \exp \left \{ \tau \hat{L}_{1} \right \} \delta(\mathbf{r}_{1}-\mathbf{r}_{0}).
\end{aligned}
\label{FF_AF3}
\end{equation}

The major difficulty in evaluating Eq.~(\ref{FF_AF3}) is imposed by the necessity to calculate the function

\begin{equation}
\tilde{W}(\mathbf{r}_{1},\mathbf{r}_{0};t) \equiv  \exp \left \{ \tau \hat{L}_{1} \right \}
  = \exp \left \{ D_{0}t \left ( \Delta_{1} + \frac{1}{2kT} \left [ \Delta_{1} U(\mathbf{r}_{1})-\frac{1}{2kT} (\nabla_{1} U(\mathbf{r}_{1}))^{2} \right ] \right ) \right \},
\label{PROP_1}
\end{equation}

which cannot be performed for the general case. Let us, therefore, consider an approximation in which the terms containing the potential energy in the operator $\hat{L}_{1}$ given by Eq.~(\ref{L1}) are neglected. Thus, we assume that $\hat{L}_{1} = \Delta_{1}$.

Under this condition, $\tilde{W}(\mathbf{r}_{1},\mathbf{r}_{0};t)$ results as the Green's function for free diffusion, i.e.

\begin{equation}
\begin{aligned}
\tilde{W}(\mathbf{r}_{1},\mathbf{r}_{0};t) & = \exp \left \{ \tau \Delta_{1} \right \} \\
& = \left ( \frac{1}{4\pi D_{0}t} \right )^{3/2} \exp \left \{ - \frac{(\mathbf{r}_{1}-\mathbf{r}_{0})^{2}}{4D_{0}t} \right \}.
\end{aligned}
\label{PROP_2}
\end{equation}

The approximation considered turns out to be reasonable only for sufficiently large pores sizes $R >> a_{0}$ (here by $R$ we denote an average pore size) and for the observation times $t$ satisfying

\begin{equation}
\tau_{R}=\frac{R^{2}}{D_{0}} >> t >> \frac{a_{0}^{2}}{D_{0}}=\tau_{0}.
\label{TimeC}
\end{equation}

In Eq.~(\ref{TimeC}), $\tau_{0}$ is the characteristic time to diffuse through the surface layer with the non-zero potential and $\tau_{R}$ is the characteristic time to diffuse over the distances comparable to the pore size $R$. Note that the behavior of $\langle \mathbf{f}(t)\mathbf{f}(0) \rangle$ is determined by only the particles which are found within the surface layer of thickness $a_{0}$ near the pore walls at $t=0$ (anywhere except this layer $\mathbf{f}(0)=0$). Thus, the condition given by Eq.\ref{TimeC} ensures, on one hand, that the statistics of the particle trajectories between two successive encounters of the surface layers is described by Eq.~(\ref{PROP_2}). On the other hand, it assures that the surface orientation at these points, where the particle encounters the surface layer, does not change appreciably.

With Eq.~(\ref{PROP_2}), the force-force autocorrelation function simplifies to

\begin{equation}
\begin{aligned}
\langle \mathbf{f}(t)\mathbf{f}(0) \rangle = \frac{(2kT)^{2}}{Z} \left ( \frac{1}{4\pi D_{0}t} \right )^{3/2} \int \int & d^{3}\mathbf{r}_{1} d^{3}\mathbf{r}_{0} \left ( \frac{\partial}{\partial \mathbf{r}_{1}} e^{-\frac{U(\mathbf{r}_{1})}{2kT}}  \frac{\partial}{\partial \mathbf{r}_{0}} e^{-\frac{U(\mathbf{r}_{0})}{2kT}} \right ) \\
& \times \exp \left \{ - \frac{(\mathbf{r}_{1}-\mathbf{r}_{0})^{2}}{4D_{0}t} \right \}.
\end{aligned}
\label{FF_AF4}
\end{equation}

Let us fix $\mathbf{r}_{0}$ and introduce a local coordinate system with the $z(\mathbf{r}_{0})$-axis being perpendicular and with the $x(\mathbf{r}_{0})$- and $y(\mathbf{r}_{0})$-axes being parallel to the closest pore wall, respectively. We may further assume that the potential $U(\mathbf{r})$ varies only along $z(\mathbf{r}_{0})$, but not along $x(\mathbf{r}_{0})$ and $y(\mathbf{r}_{0})$. For the condition given by Eq.~(\ref{TimeC}), Eq.~(\ref{FF_AF4}) can now be integrated over the coordinates of the final position vector $\mathbf{r}_{1}$ which are parallel to $x(\mathbf{r}_{0})$ and $y(\mathbf{r}_{0})$:

\begin{equation}
\begin{aligned}
\langle \mathbf{f}(t)\mathbf{f}(0) \rangle = \frac{(2kT)^{2}}{Z} \left ( \frac{1}{4\pi D_{0}t} \right )^{1/2} \int \int \int \int & d\tilde{z}_{1} d\tilde{z}_{0} d\tilde{x}_{0} d\tilde{y}_{0} \left ( \frac{\partial }{\partial \tilde{z}_{1}} e^{-\frac{U(\tilde{z}_{1})}{2kT}} \frac{\partial }{\partial \tilde{z}_{0}} e^{-\frac{U(\mathbf{r}_{0})}{2kT}}  \right ) \\
& \times \exp \left \{ - \frac{(\tilde{z}_{1}-\tilde{z}_{0})^{2}}{4D_{0}t} \right \}.
\end{aligned}
\label{FF_AF5}
\end{equation}

Note that, under the condition set by Eq.~(\ref{TimeC}), the newly introduced curvilinear coordinate system $\{ \tilde{x}_{i}(\mathbf{r_{0}}), \tilde{y}_{i}(\mathbf{r_{0}}), \tilde{z}_{i}(\mathbf{r_{0}}) \}$ is a function of only the initial position vector $\mathbf{r}_{0}$, but not of $\mathbf{r}_{1}$. Indeed, as it has been pointed out earlier, the global surface curvature in the short time approximation is neglected.

The scalar product in the brackets in the integral on the right side hand of Eq.~(\ref{FF_AF5}) is non-zero only in the vicinity of the pore walls, i.e. for $|\tilde{z}_{i}(\mathbf{r}_{0})|<a_{0}$ where $\partial U(\mathbf{r}) / \partial \mathbf{r}$ is non-zero. Hence, due to $D_{0}t>>a_{0}^{2}$, the exponential term in the integral is

\begin{equation}
\exp \left \{ - \frac{(\tilde{z}_{1}-\tilde{z}_{0})^{2}}{4D_{0}t} \right \} \approx 1.
\label{EXP1}
\end{equation}

This allows to integrate Eq.~(\ref{FF_AF5}) over $\tilde{z}_{i}$, which results in

\begin{equation}
\int d\tilde{z}_{i} \frac{\partial }{\partial \tilde{z}_{i}} e^{-\frac{U(\tilde{z}_{i})}{2kT}} = 1.
\label{EXP2}
\end{equation}

Integration over $\tilde{x}_{i}(\mathbf{r}_{0})$ and $\tilde{y}_{i}(\mathbf{r}_{0})$ yields the total surface area $S$ of the pore walls,

\begin{equation}
\int d\tilde{x}_{0}(\mathbf{r_{0}}) d\tilde{y}_{0}(\mathbf{r_{0}}) = S.
\label{EXP3}
\end{equation}

\begin{figure}
\begin{center}
\includegraphics[scale=0.33]{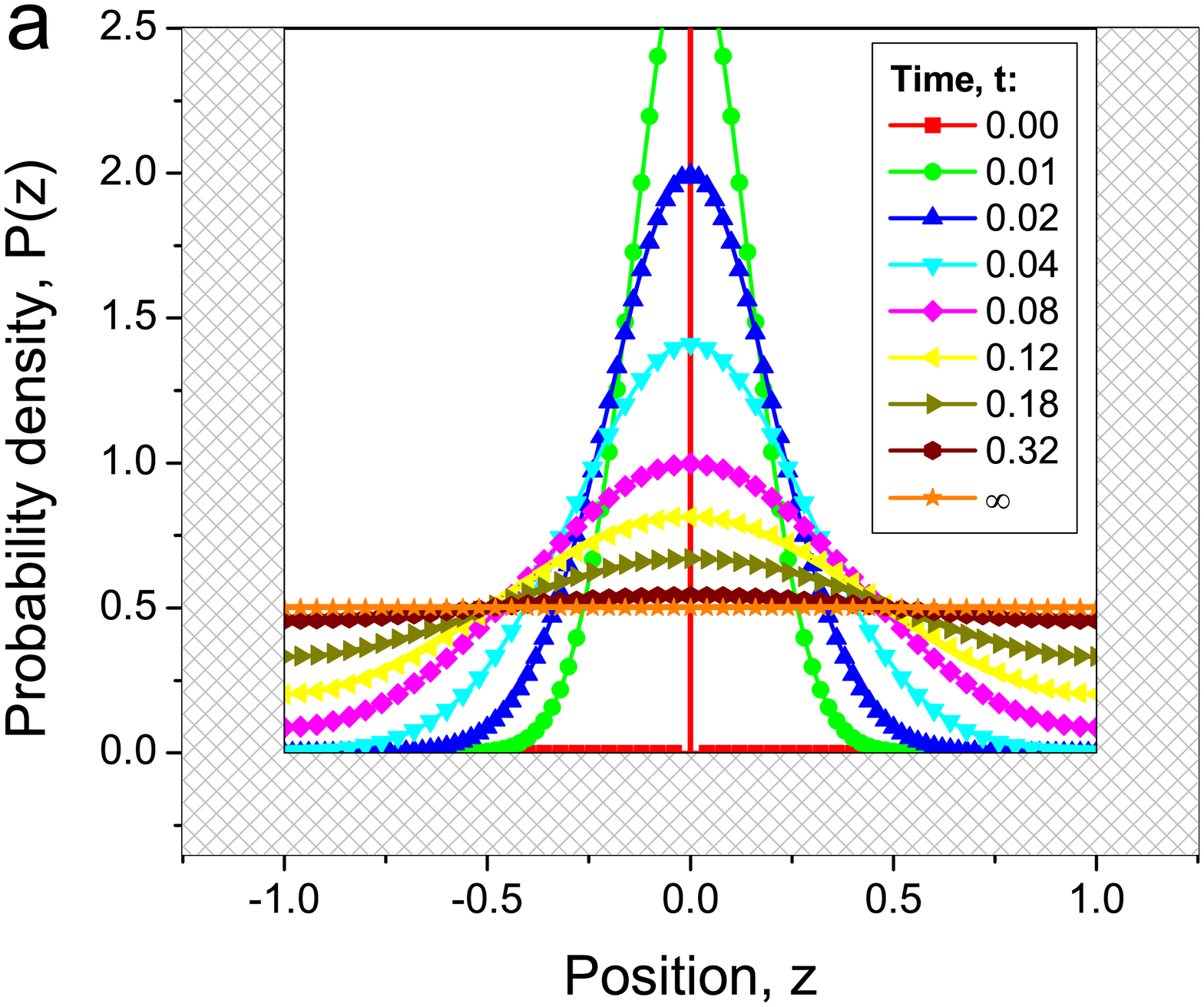}
\includegraphics[scale=0.33]{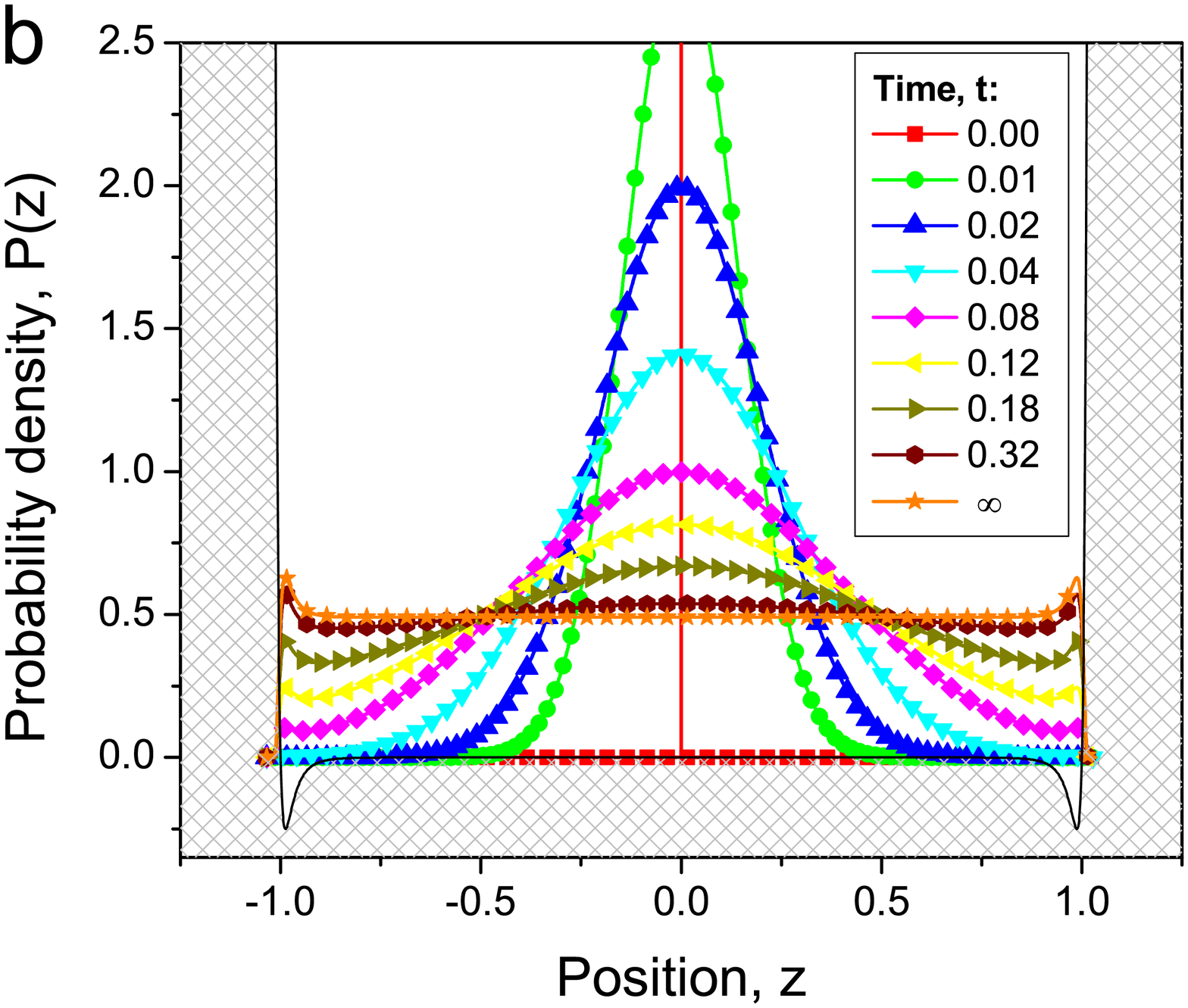}
\end{center}
\caption{The Green's functions obtained by numerical solution of the (a) Fick's and (b) Smoluchowski diffusion equation with the reflective boundary conditions at the planes located at $z=\pm 1$. The initial density distribution was $\delta (z-z_{0})$ with $z_{0}=0$. The time instances for which the solutions are shown are indicated in the inset. The black solid line in (b) shows the interaction potential used for the calculations.}
\label{Fig_FickP}
\end{figure}

\begin{figure}
\begin{center}

\includegraphics[scale=0.33]{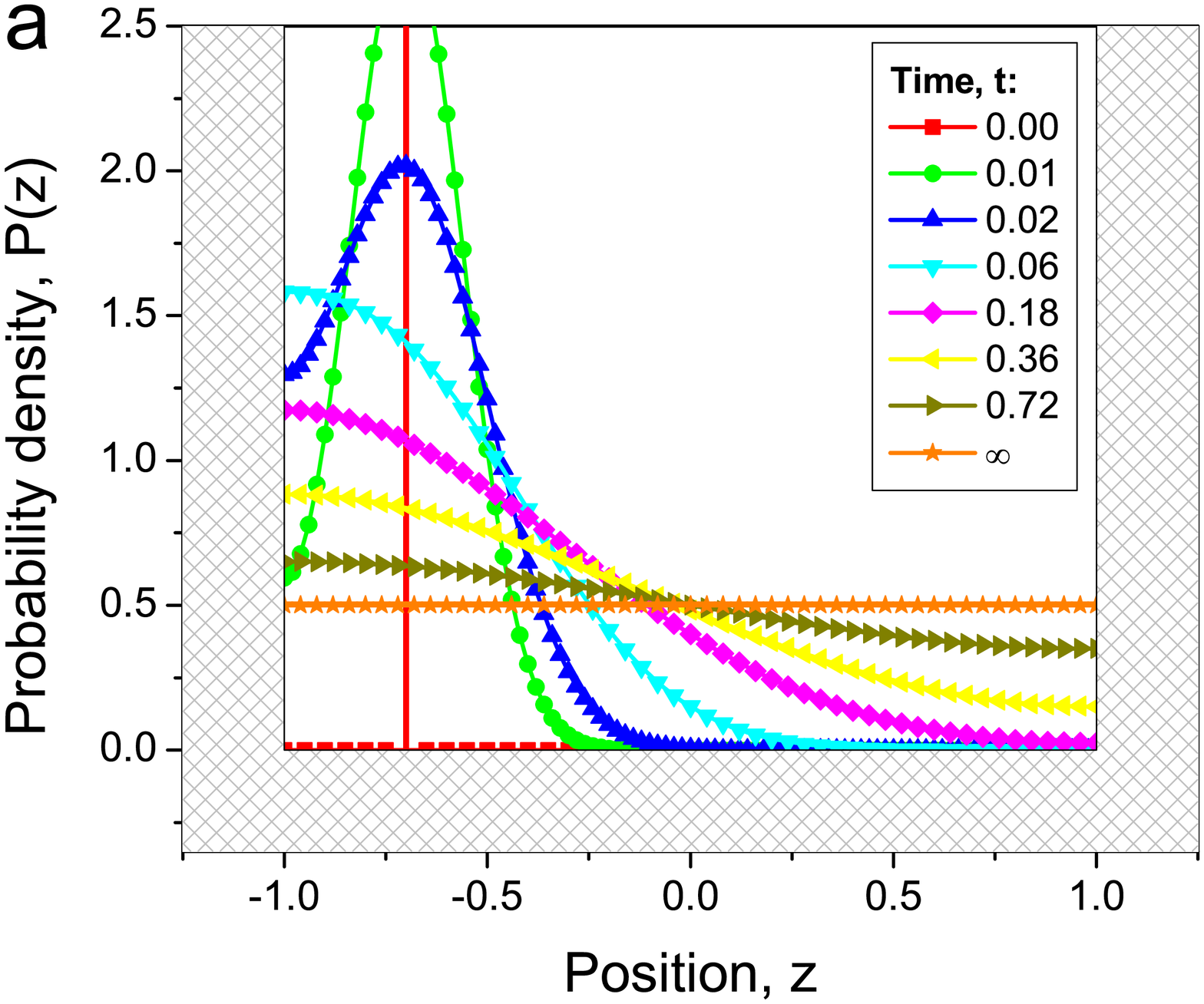}
\includegraphics[scale=0.33]{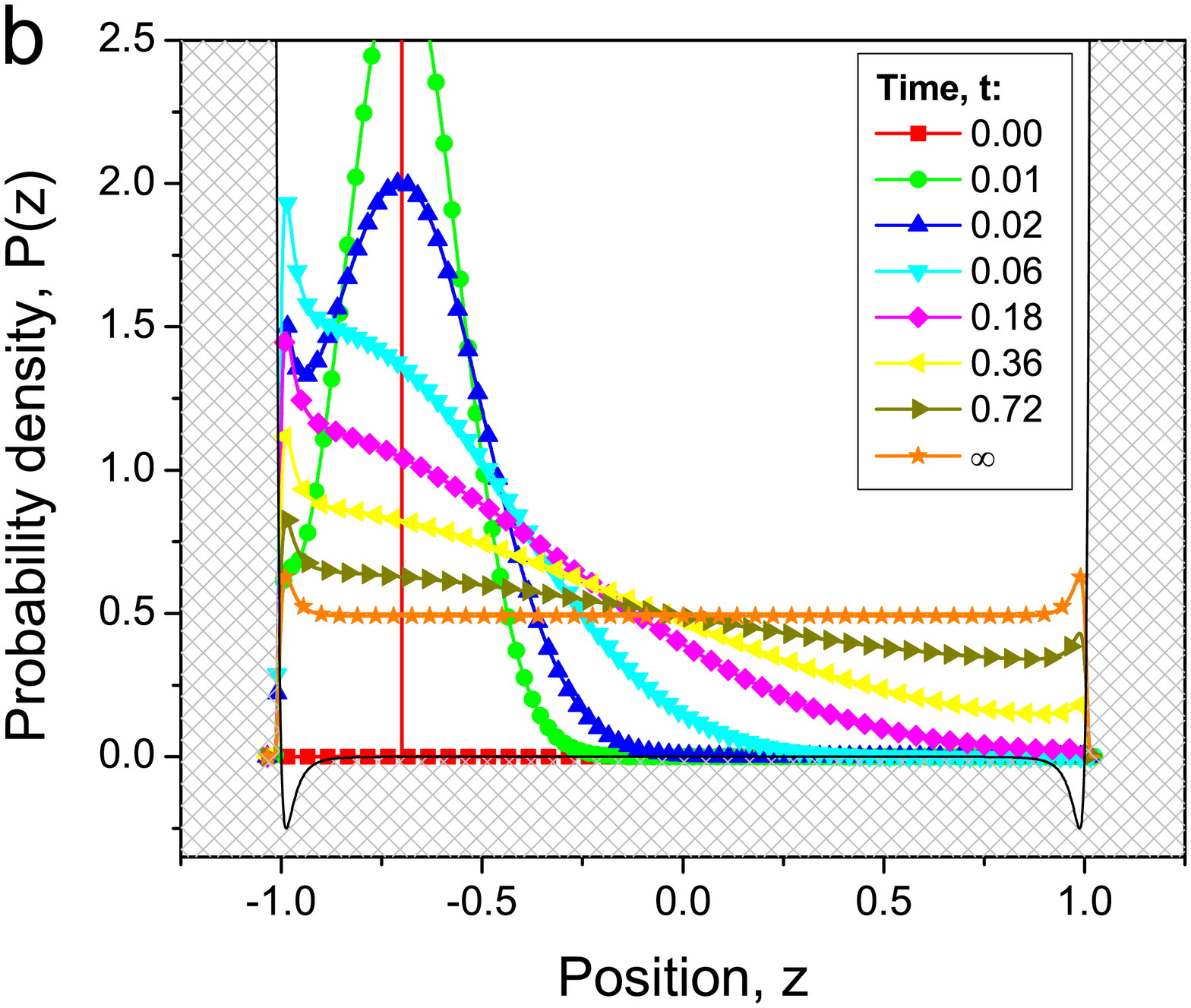}
\end{center}
\caption{The same as in Fig.~\ref{Fig_FickP}, but for $z_{0}=-0.7$.}
\label{Fig_SmolP}
\end{figure}

With Eqs.~(\ref{SSUM}), (\ref{EXP2}), and  (\ref{EXP3}), $\langle \mathbf{f}(t)\mathbf{f}(0) \rangle$  can now be readily obtained:

\begin{equation}
\langle \mathbf{f}(t)\mathbf{f}(0) \rangle =  (kT)^{2} \frac{S}{V_{p}} \frac{2}{\sqrt{\pi D_{0}t}}.
\label{FF_AFF}
\end{equation}

Finally, by substituting Eq.~(\ref{FF_AFF}) into Eq.~(\ref{DE_AF}), the short-time effective diffusivity $D_{e}$ is found to be

\begin{equation}
D_{e}(t) = D_{0} \left( 1- \frac{8}{9} \frac{S}{V_{p}} \sqrt{ \frac{D_{0}t}{\pi} } \right ).
\label{STD8}
\end{equation}

Although the functional form of this expression is compliant with that given by Eq.~(\ref{STD}) obtained earlier using the Fick's diffusion equation, they differ by a numerical constant in the term containing $\sqrt{D_{0}t}$. In order to enlighten the origin of this discrepancy, i.e. to elucidate whether this is caused by the particular approximations used or does represent the fundamental difference between the two approaches, we solved numerically both the Fick's and Smoluchowski equations with the appropriate boundary conditions. The respective results are presented in the next section.

\section{Numerical analysis}

\begin{figure}
\begin{center}
\includegraphics[scale=0.33]{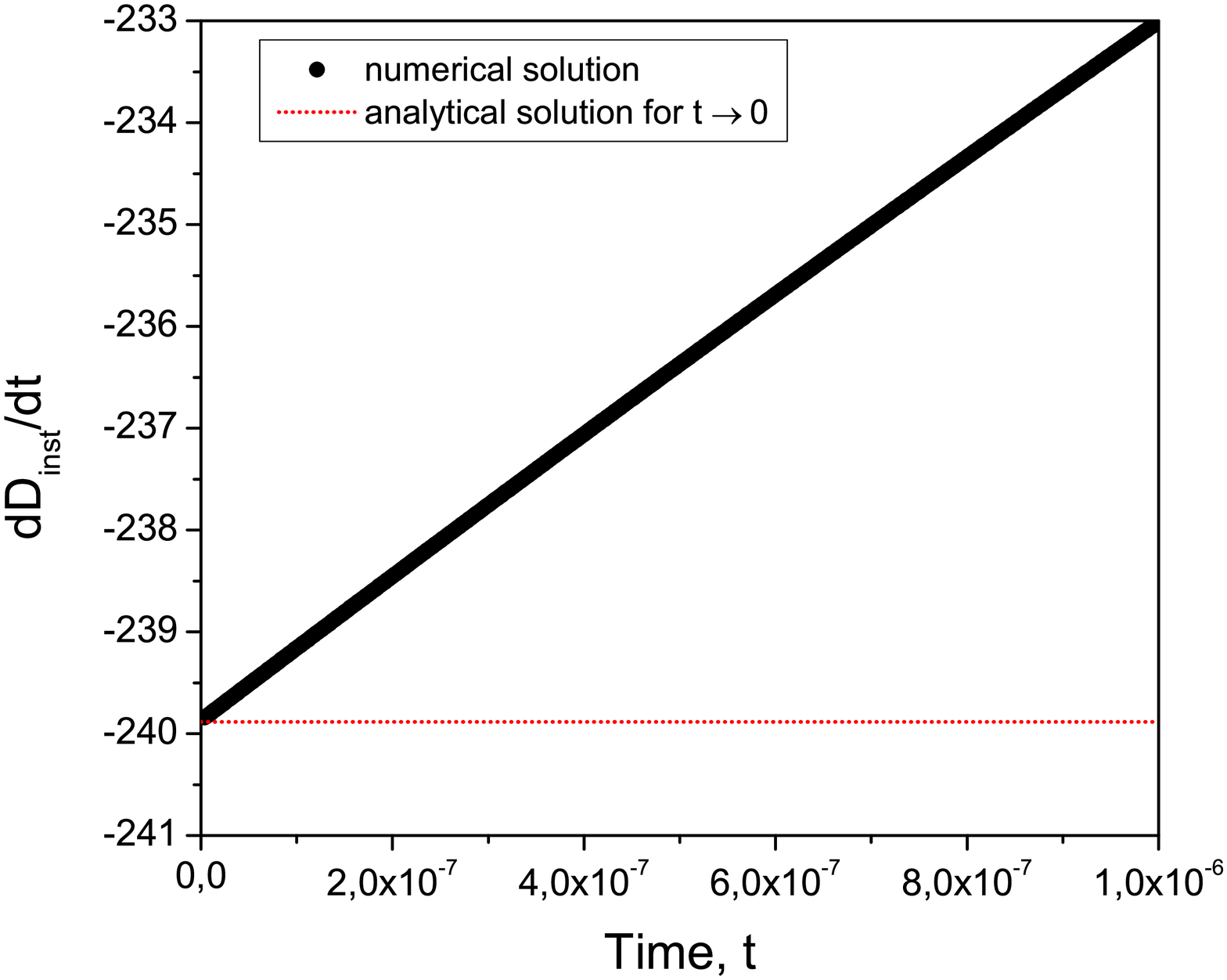}
\end{center}
\caption{Time derivative of the instantaneous diffusion coefficient $D_{inst}$ obtained by numerically solving the Smoluchowski diffusion equation with the interaction potential given by Eq.~\ref{LJ}. The horizontal line shows the theoretical prediction obtained with Eq.~\ref{DINST}.}
\label{Fig_Dinst}
\end{figure}

The numerical analysis was performed for the slit-like geometry of the pore space, i.e. diffusion occurring between two infinite planes was considered. This reduced the problem to only one spatial dimension without loss of generality. To solve the Fick's equation, reflective Neumann boundary conditions at the pore walls were applied. The Smoluchowski diffusion equation was numerically solved using several interaction potentials, which all have shown similar behavior. For the sake of conciseness, only the results obtained with the Lennard-Jones potential,

\begin{equation}
\begin{aligned}
U(x) &= 4\epsilon \left [ \frac{\sigma^{12}}{(1+\sigma+z)^{12}} - \frac{\sigma^{6}}{(1+\sigma+z)^6} \right ] \\
    & + 4\epsilon \left [ \frac{\sigma^{12}}{(1+\sigma-z)^{12}} - \frac{\sigma^{6}}{(1+\sigma-z)^6} \right ],
\end{aligned}
\label{LJ}
\end{equation}

are shown. The parameters used for the Lennard-Jones potential were $\epsilon=1/4$ and $\sigma=1/10$.

The initial conditions corresponded to the unit densities in the form of the $\delta$-functions located at certain positions in the pore space. To obtain the solution, a variable-grid implicit finite difference method was used. Some typical resulting solutions, namely the propagators, are shown in Figs.~\ref{Fig_FickP} and \ref{Fig_SmolP}. With these propagators, MSDs were obtained as the Boltzmann-weighted averages over all initial positions.

In order to validate the convergence of the numerical solutions, the limiting cases for which analytical solutions are known were used. We have considered, in particular, zero-time limit of the time derivative of the instantaneous diffusion coefficients $D_{inst}$ defined as

\begin{equation}
D_{inst} = \frac{1}{2} \frac{d \langle z^{2}(t) \rangle }{dt}.
\label{INST}
\end{equation}

The time derivative of $D_{inst}$ is just proportional to the dispersion of the external force \cite{Fatkullin1990}. This can be rationalized by recalling the definition of the force-force autocorrelation function, Eq.~(\ref{FF_AF}). For the one-dimensional case it is found to be

\begin{equation}
\frac{d D_{inst}(t)}{dt} |_{t=0} = -  \frac{D_{0}}{(kT)^{2}} \langle f^{2}(z) \rangle .
\label{DINST}
\end{equation}

The dispersion $\langle f^{2}(z) \rangle$ is readily obtained with Eq.~\ref{LJ} and can be calculated with known $\sigma$ and $\epsilon$. As demonstrated by Fig.~\ref{Fig_Dinst}, the results obtained by the numerical solutions of the Smoluchowski equation and by direct calculation for the potential field given by Eq.~\ref{LJ} are found to be in excellent agreement, validating thus the accuracy of the numerical solutions.

\begin{figure}
\begin{center}
\includegraphics[scale=0.33]{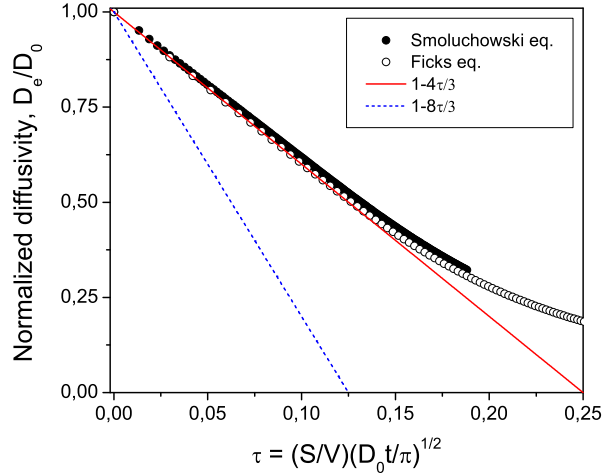}
\end{center}
\caption{Normalized diffusivity resulting as the numerical solutions of the Fick's (open circles) and Smoluchowski (filled circles) diffusion equations as a function of the parameter $\tau$. The solid and broken lines are the predictions of Eqs.~\ref{STD} and \ref{STD8}, respectively.}
\label{Fig_Deff}
\end{figure}

Finally, the diffusivities obtained for the short diffusion times using the data of Figs.~\ref{Fig_FickP} and \ref{Fig_SmolP} are shown in Fig.~\ref{Fig_Deff} as a function of the dimensionless parameter $\tau = (S/V) \sqrt{D_{0}t/\pi}$, determining the functional form of Eqs.~\ref{STD} and \ref{STD8}. Note that, for the one-dimensional case $S/V=2/L$, where $L$ is the separation between the confining planes. In the case of the confining potential (given, e.g., by Eq.~\ref{LJ}) $L^{2}$ is the dispersion of the mean square displacements for $t\rightarrow\infty$ and can be calculated exactly. As the main conclusion, the numerical analysis performed shows that the diffusivities obtained within both approaches are identical and both are described by Eq.~\ref{STD}. This reveals that the analytical result for the short-time behavior given by Eq.~\ref{STD8} contains incorrect numerical constant resulting from the approximations used upon the derivations.

\section{Discussion}

The short-time diffusion behavior, as given by Eq.~(\ref{STD}), was derived by Mitra \textit{et al.} by solving the Fick's diffusion equation \cite{Mitra1992}. The existence of the impermeable pore walls was accounted for by considering reflective boundary conditions at the pore walls, i.e. $\partial W(\mathbf{r};t)/\partial r_{n}$=0. In contrast, this work treats the confinements on a more fundamental level by introducing an interaction forces between the particles and solid. Under this condition, namely in the presence of an external, spatially-varying force field, the problem is described by the Smoluchowski diffusion equation \cite{Gillespie2012}. The penalty for using this way of modeling the confinements is the complexity of the Smoluchowski equation, which cannot be solved generally for an arbitrary potential field. Hence, the analysis shall be based on certain approximations.

As such an approximation, we derived the short-time diffusion behavior by neglecting all potential-dependent terms in the Liouville operator given by Eq.~(\ref{L1}). The result obtained in this way may be considered as a rigorous upper-bond solution for the numerical constant in the term containing $\sqrt{D_{0}t}$ in Eq.~(\ref{STD8}). This can be shown by taking into account the terms in the operator $\hat{L}_{1}$ given by Eq.~(\ref{L1}), which were neglected upon the derivation of Eq.~(\ref{STD8}). It is easy to see that the integral

\begin{equation}
\int_{-\sqrt{D_{0}t}}^{\sqrt{D_{0}t}} d\tilde{z}_{1} \left ( \frac{\partial^{2} U(z_{1})}{\partial z_{1}^{2}} - \frac{1}{2kT} \left ( \frac{\partial U(z_{1})}{\partial z_{1}} \right )^{2} \right ) \approx -\frac{1}{2kT} \int_{-\sqrt{D_{0}t}}^{\sqrt{D_{0}t}} d\tilde{z}_{1} \left ( \frac{\partial U(z_{1})}{\partial z_{1}} \right )^{2} \leq 0
\label{UB}
\end{equation}

is always negative. This means that the amplitude of the Green's function, as given by  Eq.~(\ref{PROP_1}), decreases with increasing time faster as compared to the truncated Green's function given by Eq.~(\ref{PROP_2}). The faster the Green's function attenuates, the higher is the effective diffusivity. The latter implies a lower numerical constant in Eq.~(\ref{STD8}).

It is interesting to note that the difference between the numerical pre-factors in Eqs.~\ref{STD} and \ref{STD8} is exactly the factor 2. An analogous difference has already been noted in similar contexts \cite{Buonocore1987, Schofield1992, Fuchs2002, Sturm2014}. In particular, it resulted (i) when the effect of the sharp, adsorbing boundaries on diffusion was modeled by the method of images \cite{Sturm2014} or (ii) when a statistical mechanics derivation of a hydrodynamic boundary condition for the diffusion equation was obtained \cite{Fuchs2002}. In the latter case, the omission of the higher-order density projections terms also led to a two-fold underestimate of a respective coefficient \cite{Schofield1992}. In a similar manner, the truncation of the Liouville operator in our case left it with only the term resulting in the propagator for free diffusion, Eq.~(\ref{PROP_2}). Under this approximation, the inward net flux due to the confining potential turns out to be as twice as large than that obtained with the impermeable boundaries. Thus, one may expect that accounting for the two neglected terms in the square brackets of Eq.~\ref{L1} should diminish the flux correspondingly and to result in Eq.~(\ref{STD}).

In this respect, it is instructive to see under which conditions Eq.~(\ref{STD}) can be obtained from the Smoluchowski equation. This can be done phenomenologically. Because we are concerned with the time intervals satisfying Eq.~(\ref{TimeC}), the Brownian particles can displace by the distances $\sqrt{D_{0}}t$ much smaller than the typical curvature $R$ of the pore walls (near-field approximation). Hence, diffusion parallel to the pore walls remains unaltered by the presence of the solid matrix and is described by the Green's function for free diffusion. In this case, the integration over the $\tilde{x}_{1}(\mathbf{r}_{0})$ and $\tilde{y}_{1}(\mathbf{r}_{0})$ in Eq.~(\ref{FF_AF4}) can be performed. On the other hand, the times considered are much longer than the typical times $\tau_{0}\approx a_{0}^{2}/D_{0}$, which the particles spend within the surface layers where they interact with the pore walls. This means that, during these time intervals, thermodynamic equilibrium is established. With this assumption, the functional form of the Green's function in the direction collinear with $\tilde{z}_{1}(\mathbf{r}_{0})$ can reasonably be approached by the Green's function obtained for free diffusion in a half-space convoluted with the equilibrium Boltzmann distribution, i.e.

\begin{equation}
\begin{aligned}
W_{z}(z_{1};t) = & \frac{2}{\sqrt{4\pi D_{0}t}} \exp \left \{ - \frac{z_{1}^{2}}{4D_{0}t} \right \} \\
& \times \exp \left \{ - \frac{U(z_{1})}{kT} \right \}.
\end{aligned}
\label{PROP_F}
\end{equation}

With Eqs.~(\ref{PROP_F}) and (\ref{FF_AF}), Eq.~(\ref{STD}) is readily obtained.

The condition expressed by Eq.~(\ref{TimeC}) is, thus, an important prerequisite for the establishment of the short-time diffusion regime described by Eq.~(\ref{STD}). In the opposite case, i.e. with the displacements being shorter than or of the order of $a_{0}$, the simple ansatz given by Eq.~(\ref{PROP_F}) is not validated. In this case, notably different functional forms for $D_{e}$ may be expected. Indeed, the flux reflected due to the confining potential determines the rate of change of $D_{e}$. Under continuous action of the external force, the net flux will not scale anymore with $\sqrt{D_{0}t}$, which is the property of the purely diffusive dynamics, but will be determined by the force field. The most illustrative example for this scenario may be provided by considering an one-dimensional harmonic potential $U(x)=f x^{2}/2$. The analytical solution for the Green's function for this potential is well known \cite{Doi1986} and the short-time behavior can be easily obtained:

\begin{equation}
 D_{e}(t) = D_{0} \left( 1 - \frac{\sqrt{2}}{3} \frac{f}{kT} D_{0}t \right ).
\label{Harm}
\end{equation}

This finding demonstrates the effect of the absence of the diffusive randomization, leading to a totally different short-time asymptotic.

\section{Conclusions}

Diffusion behavior of fluids in porous solids can be described by considering the confining potential induced by the pore walls on the tracer particles. This interaction potential results in an additional force exerted on the Brownian particles close to the pore walls within a surface layer of the thickness $a_{0}$. Thus, due the occurrence of the external force field one may expect some deviations from the diffusion behavior predicted for non-interacting particles near impermeable walls for short diffusion times. In this work, this problem was addressed using the respective Smoluchowski diffusion equation with the force field created by the porous solid. Only the short-time diffusion behavior, namely when the particle displacements were notably longer than $a_{0}$, but still were substantially short to neglect by the curvature of the pore walls, was analyzed. A general solution for the time-dependent diffusivity was derived, which, however, was obtained under certain approximations. Otherwise, the problem appears to be intractably complex. The thus obtained solution, Eq.~\ref{STD8}, was found to differ by a numerical constant from that obtained by solving the classical diffusion equation with reflective boundary conditions applied at the pore walls. Numerical solution of the Smoluchowski equation with selected interaction potentials allowed to identify the origin of this discrepancy, which was shown to result as a consequence of the assumptions made. It was argued that the results obtained within these two approaches should merge if, during the time intervals considered, thermodynamic equilibrium is established. The findings presented in this work may be of relevance for a broad class diffusion problems occurring in the presence of spatially-varying potential fields, such as ion diffusion near charged objects, diffusion of colloids in optical traps and viscoelastic media.

\section{Acknowledgements}

The work is supported by the German Science Foundation (DFG). We thank Klaus Kroy, University of Leipzig, for the fruitful discussions.

\section*{References}

\bibliographystyle{elsarticle-num}
\bibliography{bib_ST}

\begin{thebibliography}{10}
\expandafter\ifx\csname url\endcsname\relax
  \def\url#1{\texttt{#1}}\fi
\expandafter\ifx\csname urlprefix\endcsname\relax\def\urlprefix{URL }\fi
\expandafter\ifx\csname href\endcsname\relax
  \def\href#1#2{#2} \def\path#1{#1}\fi

\bibitem{Gillespie2012}
D.~T. Gillespie, E.~Seitaridou, Simple Brownian Diffusion: An Introduction to
  the Standard Theoretical Models, Oxford University Press, Oxford, UK, 2012.

\bibitem{Bouchaud1990}
J.~P. Bouchaud, A.~Georges, Anomalous diffusion in disordered media -
  statistical mechanisms, models and physical applications, Phys. Rep.
  195~(4-5) (1990) 127--293.

\bibitem{Bunde1996}
A.~Bunde, S.~Havlin, Fractals and Disordered Systems, 2nd Edition,
  Springer-Verlag, Berlin, Heidelberg, New York, 1996.

\bibitem{Kac1966}
M.~Kac, Can one hear shape of a drum, American Mathematical Monthly 73~(4P2)
  (1966) 1.

\bibitem{Grebenkov2007}
D.~S. Grebenkov, Nmr survey of reflected brownian motion, Rev. Mod. Phys.
  79~(3) (2007) 1077--1137.

\bibitem{Cussler2009}
E.~L. Cussler, Diffusion: Mass Transfer in Fluid Systems, 3rd Edition,
  Cambridge University Press, Cambridge, 2009.

\bibitem{Kimmich1997}
R.~Kimmich, NMR: tomography, diffusometry, relaxometry, Springer-Verlag, Berlin
  Heidelberg, 1997.

\bibitem{Price2009}
W.~S. Price, NMR Studies of Translational Motion, University Press, Cambridge,
  2009.

\bibitem{Callaghan2011}
P.~T. Callaghan, Translational Dynamics and Magnetic Resonance, Oxford
  University Press, New York, 2011.

\bibitem{Mitra1992}
P.~P. Mitra, P.~N. Sen, L.~M. Schwartz, P.~Ledoussal, Diffusion propagator as a
  probe of the structure of porous-media, Phys. Rev. Lett. 68~(24) (1992)
  3555--3558.

\bibitem{Mitra1993}
P.~P. Mitra, P.~N. Sen, L.~M. Schwartz, Short-time behavior of the
  diffusion-coefficient as a geometrical probe of porous-media, Phys. Rev. B
  47~(14) (1993) 8565--8574.

\bibitem{Latour1993}
L.~L. Latour, P.~P. Mitra, R.~L. Kleinberg, C.~H. Sotak, Time-dependent
  diffusion-coefficient of fluids in porous-media as a probe of
  surface-to-volume ratio, J. Magn. Reson. A 101~(3) (1993) 342--346.

\bibitem{Novikov2011}
D.~S. Novikov, E.~Fieremans, J.~H. Jensen, J.~A. Helpern, Random walks with
  barriers, Nat. Phys. 7~(6) (2011) 508--514.

\bibitem{Hurlimann1994}
M.~D. Hurlimann, K.~G. Helmer, L.~L. Latour, C.~H. Sotak, Restricted diffusion
  in sedimentary-rocks - determination of surface-area-to-volume ratio and
  surface relaxivity, J. Magn. Reson. A 111~(2) (1994) 169--178.

\bibitem{Sorland1997}
G.~H. Sorland, Short-time pfgste diffusion measurements, J. Magn. Reson.
  126~(1) (1997) 146--148.

\bibitem{Gjerdaker1999}
L.~Gjerdaker, G.~H. Sorland, D.~W. Aksnes, Application of the short diffusion
  time model to diffusion measurements by nmr in microporous crystallites,
  Microporous Mesoporous Mat. 32~(3) (1999) 305--310.

\bibitem{Johns2001}
M.~L. Johns, L.~F. Gladden, Surface-to-volume ratio of ganglia trapped in
  small-pore systems determined by pulsed-field gradient nuclear magnetic
  resonance, J. Colloid Interface Sci. 238~(1) (2001) 96--104.

\bibitem{Butler2002}
J.~P. Butler, R.~W. Mair, D.~Hoffmann, M.~I. Hrovat, R.~A. Rogers, G.~P.
  Topulos, R.~L. Walsworth, S.~Patz, Measuring surface-area-to-volume ratios in
  soft porous materials using laser-polarized xenon interphase exchange nuclear
  magnetic resonance, J. Phys.-Condes. Matter 14~(13) (2002) L297--L304.

\bibitem{Szutkowski2002}
K.~Szutkowski, J.~Klinowski, S.~Jurga, Nmr studies of restricted diffusion in
  lyotropic systems, Solid State Nucl. Magn. Reson. 22~(2-3) (2002) 394--408.

\bibitem{Miller2007}
G.~W. Miller, M.~Carl, J.~F. Mata, G.~D. Cates, J.~P. Mugler, Simulations of
  short-time diffusivity in lung airspaces and implications for s/v
  measurements using hyperpolarized-gas mri, IEEE Trans. Med. Imaging 26~(11)
  (2007) 1456--1463.

\bibitem{Bogdan2008}
M.~Bogdan, A.~Parnau, C.~Badea, I.~Ardelean, Time-dependent diffusion studies
  on miglyol molecules confined in permeable polymeric capsules, Appl. Magn.
  Reson. 34~(1-2) (2008) 63--69.

\bibitem{Doi1986}
M.~Doi, S.~Edwards, The Theory of Polymer Dynamics, Oxford University Press,
  Oxford, 1986.

\bibitem{Feng2007}
X.~Feng, W.~H. Thompson, Smoluchowski equation description of solute diffusion
  dynamics and time-dependent fluorescence in nanoconfined solvents, J. Phys.
  Chem. C 111~(49) (2007) 18060--18072.

\bibitem{Volpe2013}
G.~Volpe, G.~Volpe, Simulation of a brownian particle in an optical trap,
  American Journal of Physics 81~(3) (2013) 224--230.

\bibitem{Chang1975}
D.~B. Chang, R.~L. Cooper, A.~C. Young, C.~J. Martin, B.~Ancker-Johnson,
  Restricted diffusion in biophysical systems: Theory, J. Theor. Biol. 50~(2)
  (1975) 285--308.

\bibitem{Fatkullin1990}
N.~F. Fatkullin, Contribution to the theory of diffusion attenuation of
  spin-echo signals in media with random obstacles, J. Exp. Theor. Phys. 98~(6)
  (1990) 2030--2037.

\bibitem{Maklakov1992}
A.~I. Maklakov, N.~F. Fatkullin, N.~K. Dvoyashkin, Stimulated spin-echo study
  for self-diffusion of liquid molecules in random obstacle media, J. Exp.
  Theor. Phys. 101~(3) (1992) 901--912.

\bibitem{Valiullin2001}
R.~Valiullin, V.~Skirda, Time dependent self-diffusion coefficient of molecules
  in porous media, J. Chem. Phys. 114~(1) (2001) 452--458.

\bibitem{Buonocore1987}
A.~Buonocore, A.~G. Nobile, L.~M. Ricciardi, A new integral-equation for the
  evaluation of 1st-passage-time probability densities, Advances in Applied
  Probability 19~(4) (1987) 784--800.

\bibitem{Schofield1992}
J.~Schofield, I.~Oppenheim, Mode-coupling and tagged particle
  correlation-functions - the stokes-einstein law, Physica A 187~(1-2) (1992)
  210--242.

\bibitem{Fuchs2002}
M.~Fuchs, K.~Kroy, Statistical mechanics derivation of hydrodynamic boundary
  conditions: the diffusion equation, J. Phys.-Condes. Matter 14~(40) (2002)
  9223--9235.

\bibitem{Sturm2014}
S.~Sturm, J.~T. Bullerjahn, K.~Kroy, Intramolecular relaxation in dynamic force
  spectroscopy, Eur Phys J-Spec Top 223~(14) (2014) 3129--3144.

\end{thebibliography}

\end{document}